\documentclass[preprint]{aastex631}
\usepackage[utf8]{inputenc}
\usepackage{units}
\usepackage{color}
\usepackage{mathrsfs}
\usepackage{amsmath}
\usepackage{amssymb}
\usepackage{amsfonts}
\usepackage{bm}
\usepackage{natbib}
\usepackage[capitalize]{cleveref}
\usepackage{gensymb}
\usepackage{wasysym}
\usepackage[normalem]{ulem}
\hypersetup{linkcolor=red,citecolor=blue,filecolor=cyan,urlcolor=magenta}

\newcommand\phiB{\ensuremath{\phi_{\bf B}}}
\newcommand\thetaB{\ensuremath{\theta_{\bf B}}}

\shorttitle{Draft}
\shortauthors{Sabrina F. Tigik}

\begin{document}

\title{Parker Solar Probe observations of near-$f_{\rm ce}$ harmonics emissions in the near-Sun solar wind and their dependence on the magnetic field direction}

\correspondingauthor{Sabrina F. Tigik}
\email{tfsa@kth.se}

\author[0000-0002-5968-9637]{Sabrina F. Tigik}
\author[0000-0003-1654-841X]{Andris Vaivads}
\affiliation{Space and Plasma Physics, Electrical Engineering and Computer Science,\\ KTH Royal Institute of Technology, Stockholm 11428, Sweden}
\author[0000-0003-1191-1558]{David M. Malaspina}
\affiliation{Astrophysical and Planetary Sciences Department, University of Colorado, Boulder, CO, USA}
\affiliation{Laboratory for Atmospheric and Space Physics, University of Colorado, Boulder, CO, USA}
\author[0000-0002-1989-3596]{Stuart D. Bale}
\affiliation{Physics Department, University of California, Berkeley, CA 94720-7300, USA}
\affiliation{Space Sciences Laboratory, University of California, Berkeley, CA 94720-7450, USA}

\begin{abstract}
Wave emissions at frequencies near electron gyrofrequency harmonics are observed at small heliocentric distances below about 40 solar radii and are known to occur in regions with quiescent magnetic fields. We show the close connection of these waves with the large-scale properties of the magnetic field. Near electron gyrofrequency harmonics emissions occur only when the ambient magnetic field points to a narrow range of directions bounded by polar and azimuthal angular ranges in the RTN coordinate system of correspondingly $80^{\degree} \lesssim \thetaB \lesssim 100^{\degree}$ and $10^{\degree} \lesssim \phiB \lesssim 30^{\degree}$. We show that the amplitudes of wave emissions are highest when both angles are close to the center of their respective angular interval favorable to wave emissions. The intensity of wave emissions correlates with the magnetic field angular changes at both large and small time scales. Wave emissions intervals correlate with intervals of decreases in the amplitudes of broadband magnetic fluctuations at low frequencies of $10\,\unit{Hz}-100\,\unit{Hz}$. We discuss possible generation mechanisms of the waves.
\end{abstract}


\section{Introduction}
\label{sec:intro}

The NASA's Parker Solar Probe (PSP) mission was launched in August 2018 with the objective to study the near-Sun environment that has never been explored before \citep{fox2016}. Since its first encounter with the near-Sun environment, Parker Solar Probe has been measuring intense wave activity around the electron cyclotron frequency $f_{\rm ce}$ \citep{ma2021,malaspina2020,shi2022}. The study by \cite{malaspina2020} investigates the correlation between the emissions of near-$f_{\rm ce}$ waves and the angle  $(\theta_{\rm Br})$ between the ambient magnetic field vector and the instantaneous radial direction. They detect peaks in the electric field spectra, at frequencies $0.7\,f_{\rm ce} < f < 1.1\,f_{\rm ce} $, in an interval of $15\,\unit{R}_{\odot}$ (solar radii) around the closest approach to the Sun, during encounters 1 and 2. The authors find that the most favorable condition for the near $f_{\rm ce}$ waves to occur is $\theta_{\rm Br} < 25^{\degree}$, i.e., a quasi-radial ambient magnetic field. However, in a statistical analysis of the same interval, comparing the total hours of electron cyclotron harmonic emissions and the total hours of $\theta_{\rm Br} < 25^{\degree}$, the authors find that the waves are observed only during $10\% \sim 30\%$ of this time. As a result, \cite{malaspina2020} concludes that a near-radial ambient magnetic field, though necessary, is not a sufficient condition for the growth of near-$f_{\rm ce}$ waves. Earlier studies have suggested that the near-$f_{\rm ce}$ harmonics structures are composed of electron Bernstein waves, and at least two other wave modes that have not been completely identified so far \citep{malaspina2020,malaspina2021,ma2021,shi2022}. The generation mechanism of these waves is still an open question. 

In this work, we extend the analysis in \cite{malaspina2020}, focusing on the connection of the emission of near-$f_{\rm ce}$ harmonics with large-scale properties of the magnetic field. We show that near-$f_{\rm ce}$ harmonics are observed only when the ambient magnetic field points to a narrow range of directions. In addition, we show that these waves are observed only in regions where the broadband magnetic turbulence is low. The results reported in this work are crucial for understanding the origin of the near-$f_{\rm ce}$ harmonics waves measured by Parker Solar Probe at small heliocentric distances.

\section{Data}
This study uses data from PSP's $5^{\rm th}$ perihelion passage, on June 7, 2020. The closest perihelion distance in this encounter was about $\sim 27.87\,\unit{R_\odot}\,(0.13\,\unit{AU})$ at 09:06\,UTC. During this perihelion passage intense near-$f_{\rm ce}$ waves are observed under different solar wind conditions and thus it is well suited for deeper study of the waves.

We use Parker Solar Probe measurements of magnetic and electric fields from the FIELDS experiment \citep{bale2016} and ions from the SWEAP experiment \citep{kasper2016}. Magnetic field data from both the Fluxgate Magnetometer (FGM) and Search Coil Magnetometer (SCM) instruments are used. As for electric field data, we use two distinct differential electric potential data sets, both measured by the FIELDS antenna pairs $V_{12}$ and $V_{34}$: AC-coupled spectra in survey mode, with a cadence of $\sim 0.87\,\unit{s}$, and burst waveform data, sampled at $\sim 150\,\unit{kS/s}$ during $\sim 3.5\,\unit{s}$ snapshots. To convert the differential voltage data into an electric field we use $3.5\,\unit{m}$ effective antenna length \citep{mozer2020}. Search coil magnetometer data and both electric voltage data sets are produced by the Digital Fields Board (DFB) \citep{malaspina2016}. Proton velocity data are from the Solar Probe ANalyzer-Ions (SPAN-I) instrument \citep{livi2021}, which is part of the SWEAP experiment suite.

\begin{figure}[ht!]
\centering  \includegraphics[width=0.75\linewidth]{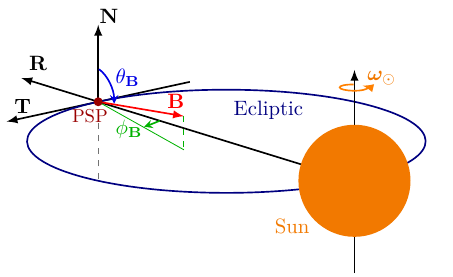}
\caption{RTN coordinate system and angles used in the study. The magnetic field vector (red arrow) represents a characteristic situation when the magnetic field is close to radial and points sunward, such as during PSP's $5^{\rm th}$ perihelion. The two angles shown are: $\theta_{\bf B}= \arccos({\rm B_N}/{\bf |B|})$ (in blue), the polar angle of ${\bf B}$ with respect to ${\bf N}$, and  $\phi_{\bf B} = \arctan({\rm B_T}/{\rm B_R})$ (in green), the azimuthal angle between the projection of the magnetic field vector onto the RT plane and the line parallel to the ${\bf R}$ direction.} \label{fig:sketch}
\end{figure}

We use two coordinate systems. In the large-scale analysis, magnetic field and proton velocity data are shown in RTN coordinates. Regarding the proton velocity, we use SPAN-I level 3 (L3), version 4 (v04) data, which includes, among other data products, ion velocities measured in SPAN-I instrument coordinates transformed to the RTN coordinate system. The ${\bf R}$ (radial) unit vector points from the Sun to the center of the spacecraft. The ${\bf T}$ (tangential) unit vector lies in the ecliptic plane and is defined as ${\bf T} = ({\bm \omega}_{\odot} \times {\bf R})/|{\bm \omega}_{\odot} \times {\bf R}|$, where ${\bm \omega}_{\odot}$ is the Sun's rotation vector. The ${\bf T}$ unit vector  points in the solar rotation direction, which is close to the PSP's ram direction during the encounters. The ${\bf N}$ (normal) unit vector completes the triad (${\bf N} = {\bf R} \times {\bf T}$) pointing upwards from the ecliptic plane. \Cref{fig:sketch} shows the representation of the RTN coordinate system. When analyzing magnetic and electric field data at small scales, we use spacecraft (SC) coordinates defined with respect to the spacecraft \citep{malaspina2016}. In the SC coordinate system, the $V_{12}$ and $V_{34}$ antennae pairs are located in the $x\, -\, y$ plane, where ${\bf \hat x}$ lies roughly in the PSP's ram direction, ${\bf \hat y}$ is approximately downwards with respect to the ecliptic plane. The third component, ${\bf \hat z}$, is perpendicular to the thermal shield, pointing towards the Sun. The SC coordinate system is optimal for making detailed comparisons between electric and magnetic fields.  

\label{sec:data}

\section{Results}
We study in detail wave emissions with frequencies around the electron gyrofrequency and its multiples. The study is divided into two parts. The first one analyses the connection of the near-$f_{\rm ce}$ waves events with large-scale properties of the ambient magnetic field. In the second part, we analyse the small scale properties of the emission intervals and the effects of low frequency waves with frequency around ion gyrofrequency on the near-$f_{\rm ce}$ waves.

\subsection{Large scale analysis}
\label{sec:large-scale-analysis}
\begin{figure}[ht!]
\centering  \includegraphics[width=\linewidth]{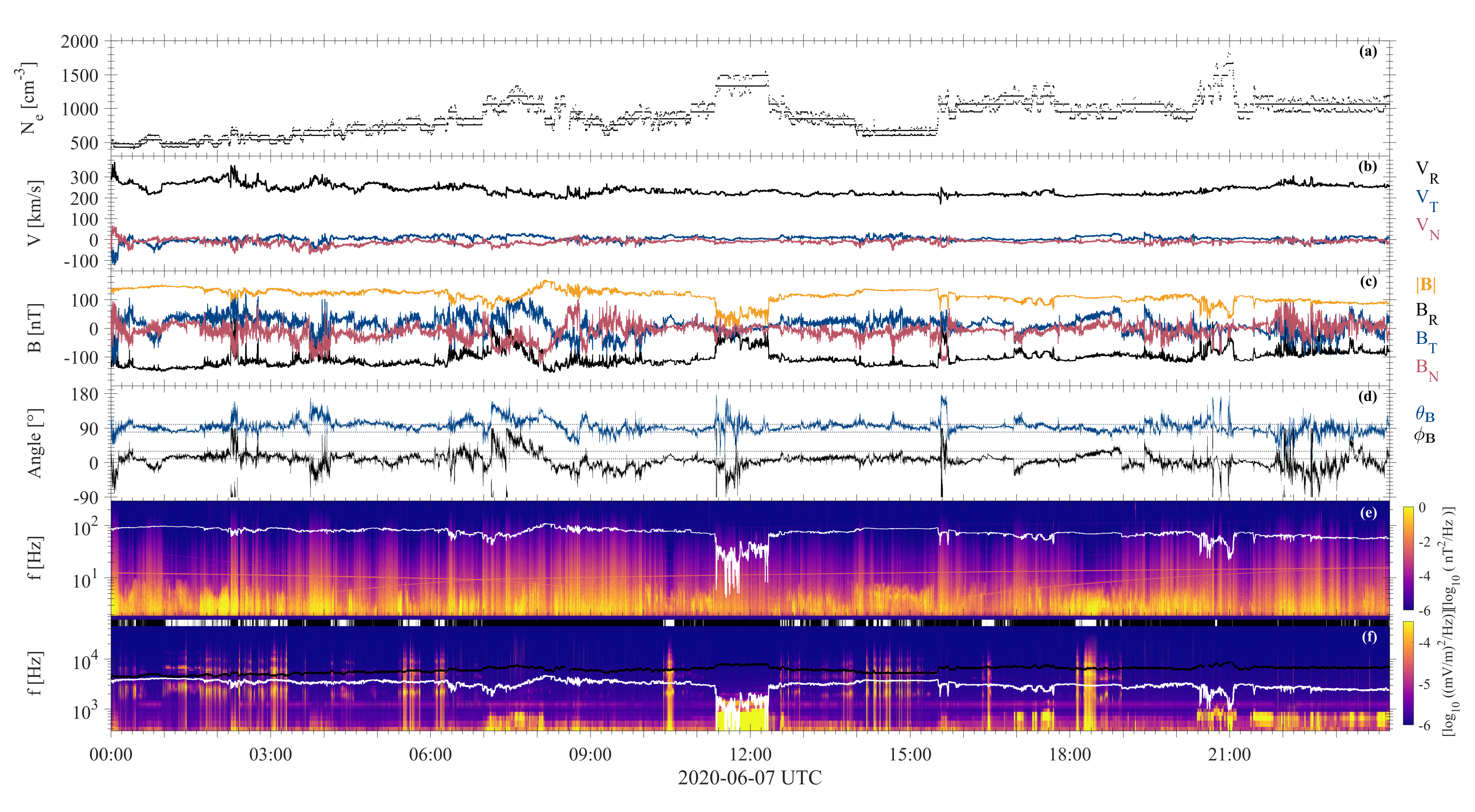}
\caption{Overview of the day surrounding the $5^{\rm th}$ perihelion of PSP. (a) Plasma density based on plasma frequency measurements. (b) Proton velocity. (c) Magnetic field. (d) The direction of magnetic field, spherical angles $\theta_{\bf B}$ and $\phi_{\bf B}$. The gray dashed lines mark the boundaries of the intervals favorable to near-$f_{\rm ce}$ harmonics emissions: $80^{\degree} < \theta_{\bf B} < 100^{\degree}$ and $10^{\degree} < \phi_{\bf B} < 30^{\degree}$. (e) Magnetic fluctuations spectrum. The white line represents the lower-hybrid frequency $f_{\rm LH}$. (f) Sum of the high-frequency AC electric field spectrum calculated from differential voltage data measured by ${\rm V}_{12}$ and ${\rm V}_{34}$ antennae pairs. The black line shows the proton plasma frequency $f_{\rm p}$ and the white line represents the electron cyclotron frequency $f_{\rm ce}$. The white markers over the black background on the top of panel (f) mark the points in time where both $\theta_{\bf B}$ and $\phi_{\bf B}$ are inside their respective angular ranges favorable to near-$f_{\rm ce}$ harmonics emissions as seen in panel (d).}
\label{fig:enc5}
\end{figure}

\Cref{fig:enc5} shows a full-day overview of PSP's $5^{\rm th}$ perihelion, on 2020 June 7. \Cref{fig:enc5}(a) depicts the plasma density from quasi-thermal noise measurements \citep{moncuquet2020}. \Cref{fig:enc5}(b) shows proton velocity from the SPAN-I instrument. During the entire day, there is a slow solar wind with the proton radial velocity varying around $250\,\unit{km/s}$. \Cref{fig:enc5}(c) shows the magnetic field. Though the magnetic field varies considerably during the day, it does not change its radial direction, with ${\rm B_R}$ pointing towards the Sun (${\rm B_R}<0$) during the whole period. \Cref{fig:enc5}(d) shows the angles $\theta_{\bf B}$ and $\phi_{\bf B}$ as defined in \Cref{fig:sketch}. Throughout the interval $\theta_{\bf B}$ mainly varies around $90^{\degree}$ and $\phi_{\bf B}$ varies around $0^{\degree}$, corresponding to a near radial magnetic field. \Cref{fig:enc5}(e) shows the magnetic field spectrogram in the frequency range from $2\,\unit{Hz}$ up to $292.7\,\unit{Hz}$. We can see broadband wave emissions with varying amplitude throughout the interval. The lower-hybrid frequency is highlighted by the white line. Finally, \Cref{fig:enc5}(f) shows the sum of the electric field spectra measured by the $V_{12}$ and $V_{34}$ antennae pairs, in a frequency range from $366\,\unit{Hz}$ to $72\,\unit{kHz}$, where the electron gyrofrequency is given by the white line and the proton plasma frequency is represented by the black line. High frequency wave emissions around the electron gyrofrequency and its multiples can be seen throughout the day; those emissions are the focus of this study.

A closer inspection of \Cref{fig:enc5} reveals that the emission of near-$f_{\rm ce}$ harmonics waves tends to occur when the background magnetic field assumes a preferential direction. Most of near-$f_{\rm ce}$ harmonics emissions coincide with instances where the magnetic field direction points to a direction within the following angle intervals $80^{\degree} \lesssim \theta_{\bf B} \lesssim 100^{\degree}$ and $10^{\degree} \lesssim \phi_{\bf B} \lesssim 30^{\degree}$. In \Cref{fig:enc5}(d) the upper and lower limits of both angle intervals are indicated by dotted gray lines. Furthermore, to illustrate this correlation, every instance the magnetic field vector is inside the region limited by the angle ranges described above is plotted in white over a black background on the top of \Cref{fig:enc5}(f). It can be noticed that near-$f_{\rm ce}$ emissions occur mostly at the same time as the white markers on top, suggesting a correlation between the waves and a particular spatial configuration of the magnetic field.

\begin{figure}[ht!]
  \centering  \includegraphics[width=0.75\linewidth]{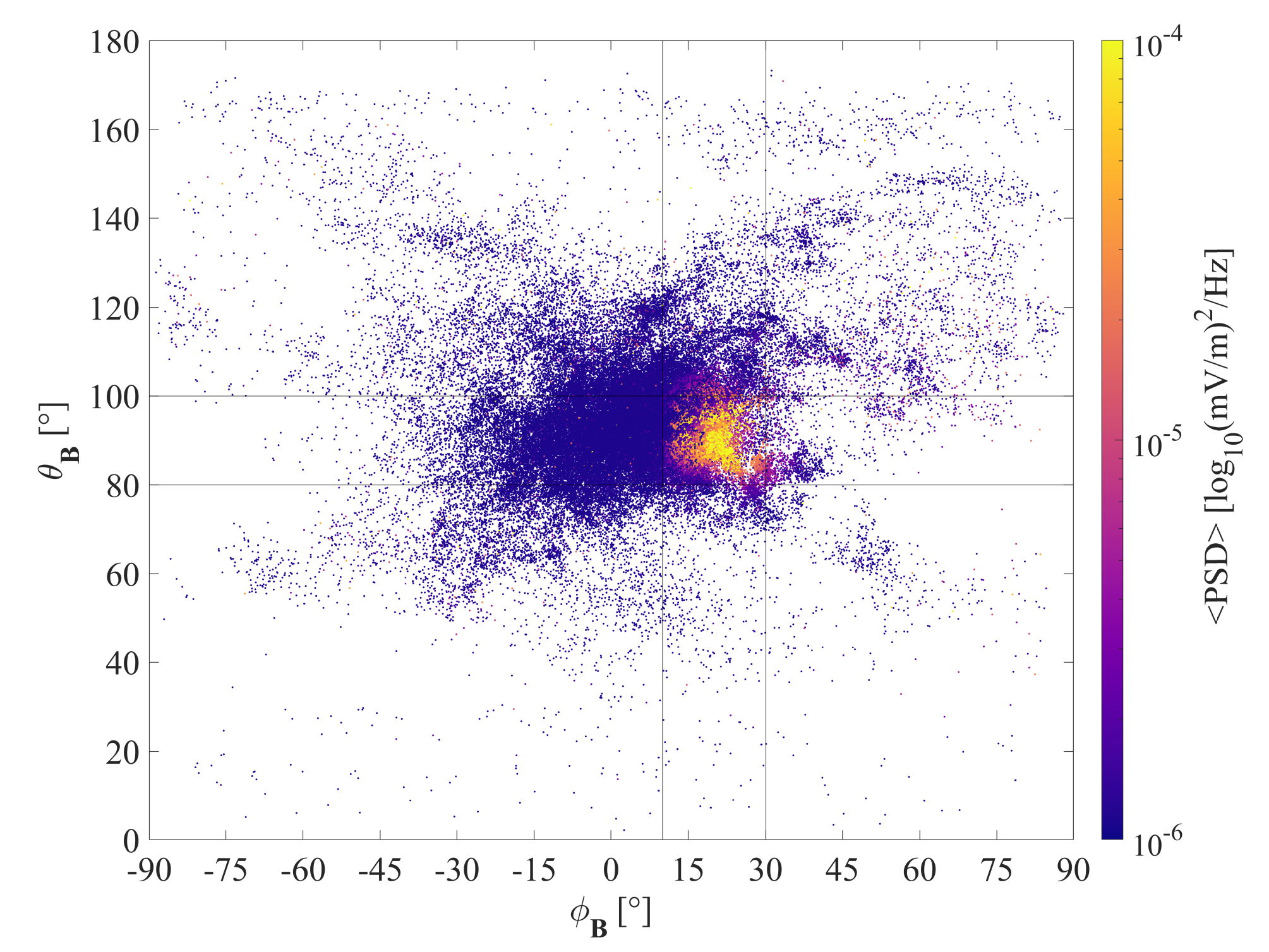}
  \caption{The average intensity of near-$f_{\rm ce}$ harmonics emissions as a function of the magnetic field direction. Data represent the whole day of 2020-06-07. Each dot corresponds to one spectral measurement averaged in the frequency range of the near-$f_{\rm ce}$ harmonics waves, $1.8\,\unit{kHz} < f < 10\,\unit{kHz}$. The magnetic field direction is characterized by  $\theta_{\bf B}$ (vertical axis) and $\phi_{\bf B}$ (horizontal axis).}
\label{fig:scatter-plot}
\end{figure}

In \Cref{fig:scatter-plot}, we further analyze whether the emission of near-$f_{\rm ce}$ harmonics indeed depends on the magnetic field direction. \Cref{fig:scatter-plot} shows the average wave power of the near-$f_{\rm ce}$ emissions  as a function of $\theta_{\bf B}$ and $\phi_{\bf B}$. The average wave power is approximated by integrating each wave spectrum in the frequency range $1.8\,\unit{kHz} < f < 10\,\unit{kHz}$. In terms of the electron gyrofrequency, the chosen frequency range corresponds to $0.55 < f/f_{\rm ce} < 3.5$, which is the optimal frequency interval for the near-$f_{\rm ce}$ harmonics during the day of the $5^{\rm th}$ perihelion, determined by visual inspection. Looking at \Cref{fig:enc5}(f), one can see that the selected frequency band is dominated by the near-$f_{\rm ce}$ harmonic waves. During a few intervals there are other type of emissions within the same frequency band, for example, in the period between 07:30 $\sim$ 9:30 there are emissions that are most probably ion-sound waves. These other waves do not seem to correlate with the magnetic field direction and are, therefore, the most probable source of the few high amplitude points randomly distributed around the scatter plot. In \Cref{fig:scatter-plot}(a) the angle range, $80^{\degree} \leqslant \theta_{\bf B} \leqslant 100^{\degree}$ and $10^{\degree} \leqslant \phi_{\bf B} \leqslant 30^{\degree}$, are marked by gray lines. One can notice that nearly all the wave power above $10^{-5}\,\unit{(mV/m)^2Hz^{-1}}$ is bounded by lines identifying the angle intervals discussed above. This confirms that the emissions of electron cyclotron harmonics peak in a limited range of magnetic field polar and azimuthal angles.

\begin{figure}[ht!]
\centering  
\includegraphics[width=0.75\linewidth]{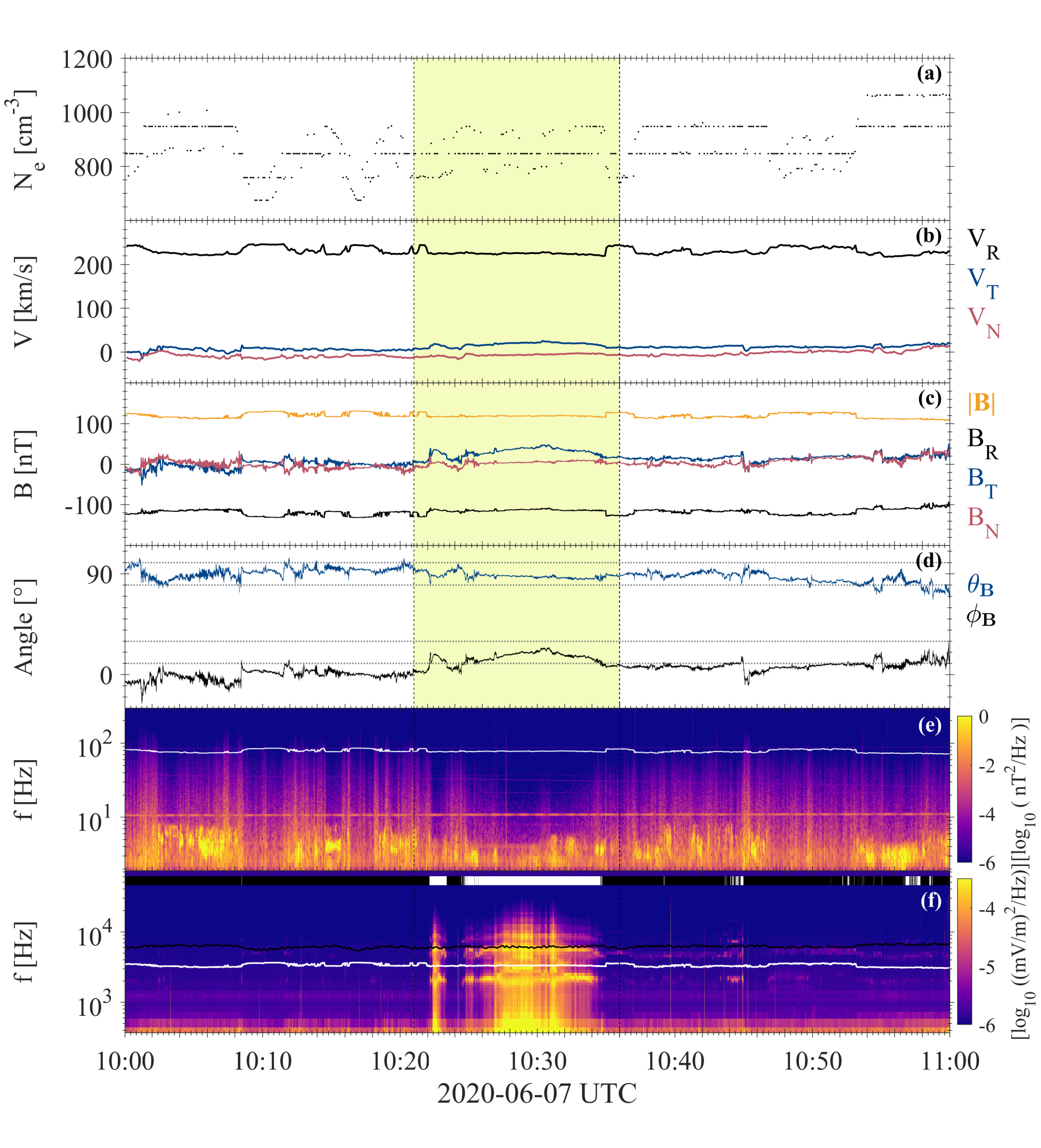}
\caption{Near-$f_{\rm ce}$ harmonics event showing a strong correlation between wave emission and variations in the magnetic field direction. Same panels as in \Cref{fig:enc5}, see detailed description there. (a) Plasma density. (b) Solar wind velocity. (c) Magnetic field. (d) Spherical angles. (e) Magnetic fluctuations. (f) AC electric field spectrum.}
\label{fig:enc5-10am}
\end{figure}

\begin{figure}[ht!]
\centering  \includegraphics[width=0.75\linewidth]{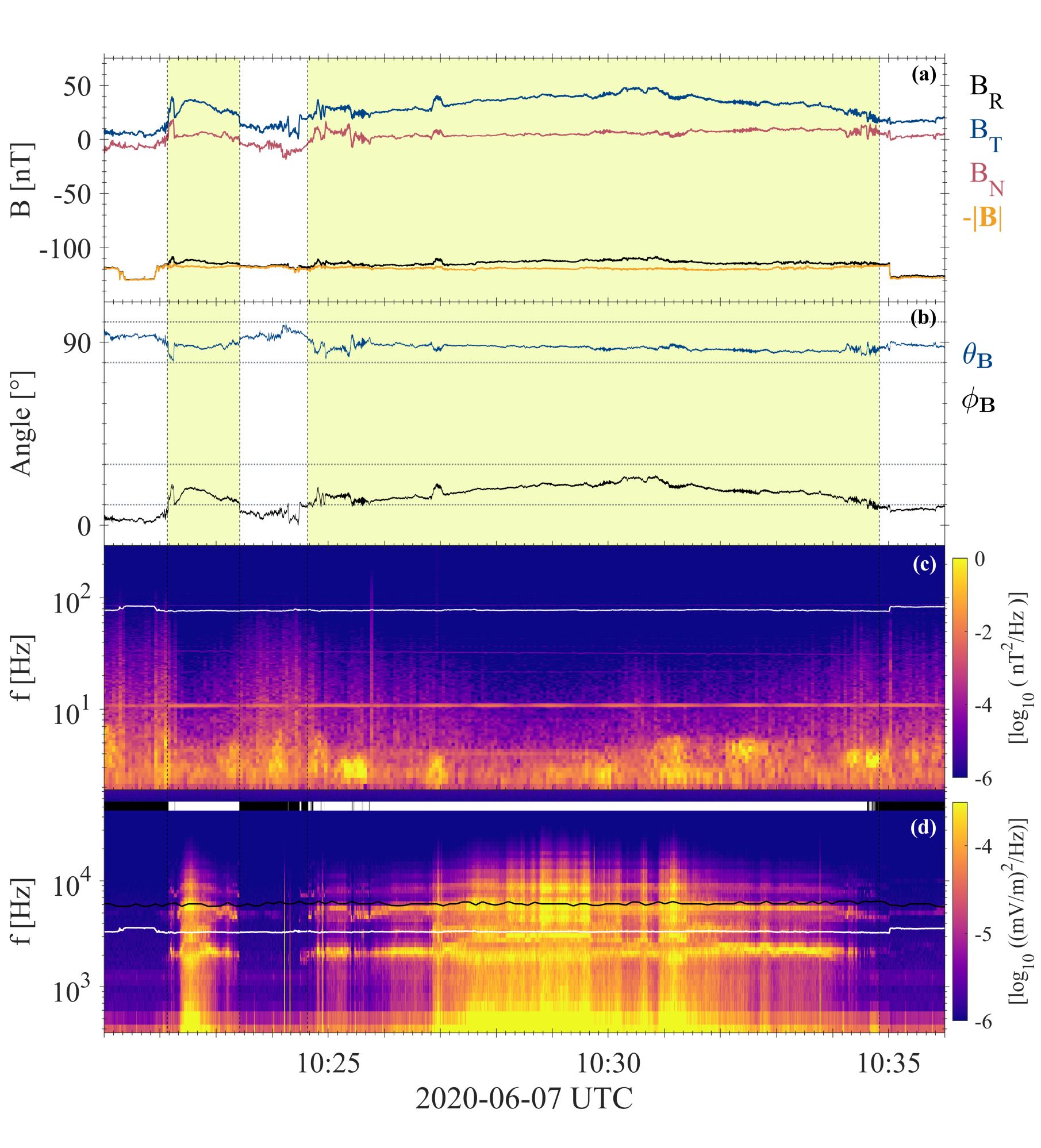}
\caption{Zoomed in figure of near-$f_{\rm ce}$ harmonics event in \Cref{fig:enc5-10am}. Shown are only the bottom four panels of \Cref{fig:enc5-10am}. (a) Magnetic field, (b) spherical angles, (c) magnetic fluctuation spectrum, and (d) AC electric field spectrum.}
\label{fig:enc5-10am-zoom}
\end{figure}

To illustrate the angular dependence of near-$f_{\rm ce}$ emissions in more detail, we select for closer analysis a one-hour interval during $5^{\rm th}$ encounter that contains a short interval of about 15min containing near-$f_{\rm ce}$ wave emissions. \Cref{fig:enc5-10am} shows the one-hour interval describing the plasma environment surrounding the wave emission event under scrutiny. Due to the stable magnetic field and very low wave activity in the electric field spectrum, where there are no other kinds of waves rather than the near-$f_{\rm cs}$ harmonics, this event provides a close to ideal situation for this analysis. Further, \Cref{fig:enc5-10am-zoom} zooms to the interval of wave emissions and we make a detailed analysis of how near-$f_{\rm ce}$ harmonic waves correlate with changes in the magnetic field.

\Cref{fig:enc5-10am}(a) and (b) show the plasma density and velocity, respectively. These two quantities do not have a determining influence on the emission of near-$f_{\rm ce}$ harmonics, but provide the large scale properties of the solar wind. \Cref{fig:enc5-10am}(c) shows the ambient magnetic field. During this time interval, the magnetic field is quite stable and mostly radial. \Cref{fig:enc5-10am}(d) depicts the spherical angles $\theta_{\bf B}$ and $\phi_{\bf B}$. The dashed gray lines mark the upper and lower boundaries of the angle intervals where we expect wave emissions to occur. \Cref{fig:enc5-10am}(e) shows the low frequency magnetic fluctuation spectrum. The white line represents the lower-hybrid frequency, and the straight line at $\sim 10\,\unit{Hz}$ is an instrumental artifact caused by the magnetic signature of the reaction wheels and has no influence on this analysis. \Cref{fig:enc5-10am}(f) depicts the high frequency electric field spectrum. The sole wave activity at high frequencies during the whole one-hour interval is the near-$f_{\rm ce}$ harmonic wave emissions. The emissions are observed only during the time intervals when both angles in \Cref{fig:enc5-10am}(d) are concurrently inside their respective angle interval. These occurrences are also marked in white over a black background on the top of \Cref{fig:enc5-10am}(f). In addition, we can see that there is a decrease in the amplitude of low frequency magnetic fluctuations during the intervals of the near-$f_{\rm ce}$ harmonic wave emissions.

\Cref{fig:enc5-10am-zoom} shows a further 15 min zoom-in on the highlighted interval in \Cref{fig:enc5-10am}. \Cref{fig:enc5-10am-zoom}(a) shows the magnetic field and \Cref{fig:enc5-10am-zoom}(b) shows the corresponding polar and azimuthal angles. The boundaries that delimit the two angular intervals favorable for wave emissions are marked with dotted lines. Periods where both $\theta_{\bf B}$ and $\phi_{\bf B}$ are inside their respective angular boundaries,  satisfying the favorable conditions for near-$f_{\rm ce}$ harmonics emission, are highlighted in yellow. Along these 15 minutes, the magnetic field is very close to the radial direction, as can be seen, by the radial component ${\rm B_R}$ being dominant and having its value very close to the negative of the magnitude of magnetic field $-{\bf |B|}$. The largest deviations from the radial direction occur in regions where ${\rm B_T}$ reach larger values. During most of the time, the normal component ${\rm B_N}$ remains relatively close to ${\rm B_N}\approx 0\,\unit{nT}$, corresponding to $\theta_{\bf B}\approx 90^{\degree}$, and throughout the whole period $\theta_{\bf B}$ stays within the interval favorable for the emissions, including the few instances where ${\rm B_N}$'s amplitude peaks at $|{\rm B_N}|\lesssim 20\,\unit{nT}$. On the other hand, $\phi_{\bf B}$, which corresponds to the ${\rm B_T}$ variations, enters and exits for several times the $\phi_{\bf B}$ interval favorable for the emissions. Thus, for this particular event, the emission of near-$f_{\rm ce}$ harmonics waves depend only on ${\rm B_T}$ variations. Looking at \Cref{fig:enc5-10am-zoom}(d), it is clear that near-$f_{\rm ce}$ harmonics emission coincide with the regions marked in yellow, where both $\phi_{\bf B}$ and $\theta_{\bf B}$ are inside their respective angular boundaries. Furthermore, the variation in the waves amplitude in \Cref{fig:enc5-10am-zoom}(d) relates to the variation in the angles within their respective intervals in \Cref{fig:enc5-10am-zoom}(b). The closer both angles are to the center of their interval favoring emissions, the stronger the wave emissions are. Such correlation could already be inferred from \Cref{fig:scatter-plot}, where one can see that the highest wave amplitudes lie close to the center of the bounded interval. However, \Cref{fig:enc5-10am-zoom} shows in detail how even a slight change in one of the spherical angles immediately impacts the properties of near-$f_{\rm ce}$ harmonics waves.

\Cref{fig:enc5-10am-zoom}(c) shows the low frequency magnetic fluctuations spectrum. The spectrum contains broadband fluctuations with varying intensity throughout the interval and, in addition, there are localized emissions having frequency peaks up to $f \approx 7\,\unit{Hz}$. Broadband magnetic fluctuations have distinctively lower amplitudes during the intervals of the near-$f_{\rm ce}$ harmonics wave emissions. The amplitude decrease follows the same correlation as described above for the near-$f_{\rm ce}$ harmonics, with the lowest broadband amplitudes corresponding to instances where both $\phi_{\bf B}$ and $\theta_{\bf B}$ are closer to the center of their respective intervals favoring wave emissions. On the other hand, the localized emissions at a few Hz do not show any direct correlation with the near-$f_{\rm ce}$ harmonics waves or the spherical angles.

\begin{figure}[ht!]
\centering  \includegraphics[width=0.75\linewidth]{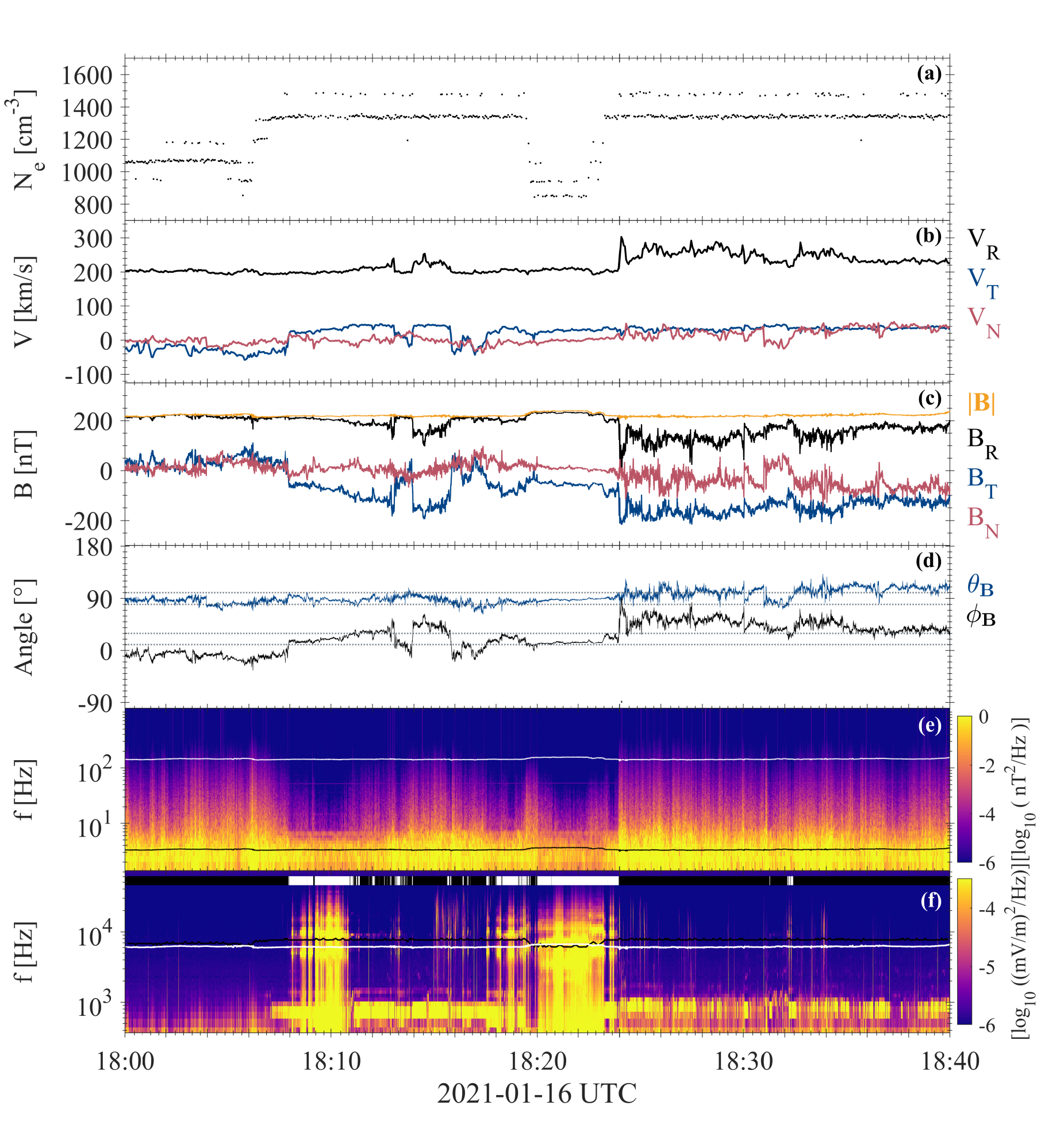}
\caption{Near-$f_{\rm ce}$ harmonics event from 2021 January 16 (orbit 7), showing the same correlations between the magnetic field direction and the high frequency electric field wave emissions. (a) Plasma density. (b) Solar wind velocity. (c) Magnetic field. (d) Spherical angles. (e) Magnetic fluctuations. (f) AC electric field spectrum.}
\label{fig:enc7-1800-1840}
\end{figure}

\Cref{fig:enc7-1800-1840} shows an interval from orbit 7, measured at a distance of $\sim 22.78\,\unit{R_{\odot}}\,(\sim 0.10\,\unit{AU})$ from the Sun. In this event, a series of near-$f_{\rm ce}$ harmonics waves occur in a more active plasma environment. The figure clearly shows that the correlations and conditions for near-$f_{\rm ce}$ harmonics wave emissions shown in \Cref{fig:enc5-10am-zoom} hold also for different solar wind conditions and different PSP distances from the Sun. \Cref{fig:enc7-1800-1840}(a) and \Cref{fig:enc7-1800-1840}(b) show the plasma density and velocity, respectively. While the plasma density is higher in this event, the slow solar wind has a similar plasma velocity as seen in \Cref{fig:enc5-10am}(b). The plasma density makes several step-wise variations, however, there is no clear correlation between the emission of near-$f_{\rm ce}$ harmonics in \Cref{fig:enc7-1800-1840}(f) and plasma density variations. \Cref{fig:enc7-1800-1840}(c) shows the magnetic field, the radial component points outwards from the Sun $({\rm B_R}>0)$. The magnetic field shows a higher degree of variations than the period shown in \Cref{fig:enc5-10am}(c). \Cref{fig:enc7-1800-1840}(d) shows the spherical angles $\phi_{\bf B}$ and $\theta_{\bf B}$. Though $\phi_{\bf B}$ varies in a larger range than $\theta_{\bf B}$, both spherical angles go in and out of their wave emission intervals. Instances where both $\phi_{\bf B}$ and $\theta_{\bf B}$ are concurrently within their respective angular boundaries correlate very well with near-$f_{\rm ce}$ harmonics emissions in \Cref{fig:enc7-1800-1840}(f). In \Cref{fig:enc7-1800-1840}(e), the low frequency magnetic fluctuations spectrum shows a clear correlation between decreased amplitudes in the broadband fluctuations and time intervals where both $\phi_{\bf B}$ and $\theta_{\bf B}$ are inside their respective boundaries. Throughout the 40 min interval, there are high amplitude wave emissions with spectra peaks at a few Hz that is close to the ion-cyclotron frequency (black line).

\subsection{Small scale analysis}
\label{sec:small-scale-analysis}

\begin{figure}[ht!]
 \centering  \includegraphics[width=0.75\linewidth]{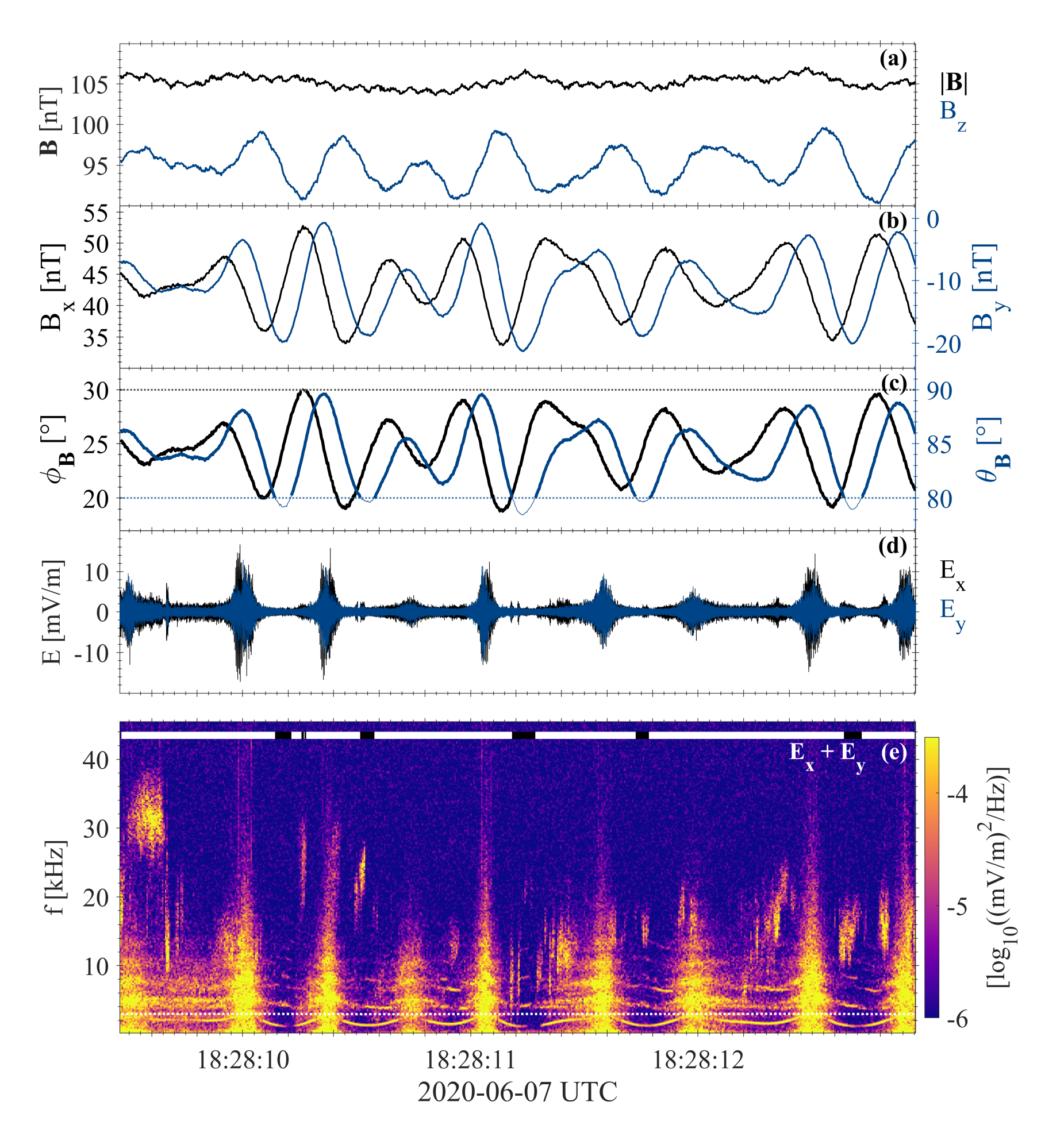}
 \caption{High-cadence $\sim 3.5\,\unit{s}$ burst data showing a series of time-localized near-$f_{\rm ce}$ harmonics wave emissions correlated with low frequency magnetic field oscillations. (a,b) Magnetic field in spacecraft coordinates. (a) ${\rm B_z}$ and $|{\rm B_z}|$. (b) ${\rm B_x}$ (left axis) and ${\rm B_y}$ (right axis). (c) Spherical angles: $\phi_{\bf B}$ (left axis) with the black dotted line marking the $\phi_{\bf B} = 30^{\degree}$ boundary, and $\theta_{\bf B}$ (right axis) with the blue dotted line marking the $\theta_{\bf B} = 80^{\degree}$ boundary. (d) Electric field waveform of the burst data. The two available components, ${\rm E_x}$ and ${\rm E_y}$, are in spacecraft coordinates. (e) Total electric field power spectrum. The white dotted line represents the electron cyclotron frequency.}
\label{fig:enc5-burst62}
\end{figure}

\Cref{fig:enc5-burst62} shows a $\sim 3.5\,\unit{s}$ near-$f_{\rm ce}$ harmonics event recorded in burst mode at $18:28:09.459$ on 2020 June 07 (PSP's $5^{\rm th}$ perihelion). Figures \ref{fig:enc5-burst62}(a) and (b) show magnetic field and high amplitude oscillations with frequencies close to the ion gyrofrequency can be seen. These are circularly polarized waves as can be seen from the $90^{\degree}$ phase shift between  ${\rm B_x}$ and ${\rm B_y}$ components.  \Cref{fig:enc5-burst62}(c) shows the azimuthal angle $\phi_{\bf B}$ and the polar angle $\theta_{\bf B}$ with the dotted lines marking the corresponding boundary of the angular range favoring wave emissions.  While $\phi_{\bf B}$ stays all the time within the angle range favorable for the wave emissions, $\theta_{\bf B}$ crosses the boundary for short intervals multiple times.  \Cref{fig:enc5-burst62}(d) shows the electric field waveform of the burst data. One can observe time-localized wave packets with amplitudes that correlate very well with the ion cyclotron wave oscillations in the magnetic field. On close inspection, we can see that minimum in the amplitude correspond to the intervals when $\theta_{\bf B}$ crosses the boundary of the angle range favorable for the wave emissions and the maximum is when $\theta_{\bf B}$ is furthest from the boundary.  \Cref{fig:enc5-burst62}(e) shows the total electric field power spectrum, where a series of time-localized near-$f_{\rm ce}$ harmonics structures, regularly distributed throughout the interval can be seen. They correspond to the largest amplitude wave packets seen in \Cref{fig:enc5-burst62}(e). In addition, besides the near-$f_{\rm ce}$ harmonics one can also distinguish localized wave emissions having peak frequencies at about a few times the electron cyclotron frequency that do not show harmonic structure at electron cyclotron frequencies. Those are most probably ion sound waves and they do not show any correlation with the magnetic field direction. In addition, there are  during the gap intervals the are narrow banded emissions that seem to connect the harmonics structures. 

\label{sec:results}

\section{Discussion}
Near-$f_{\rm ce}$ harmonics waves are observed exclusively during time intervals where the magnetic field direction is within (or very close to) the following ranges: $80^{\degree} \lesssim \thetaB \lesssim 100^{\degree}$ and  $10^{\degree} \lesssim \phiB \lesssim 30^{\degree}$ (see \Cref{fig:scatter-plot}). This corresponds to the magnetic field orientation being close to radial but slightly tilted in the direction of positive \phiB\  (${\rm B_T} > 0$ when ${\rm B_R} < 0$, or ${\rm B_T} < 0$ when ${\rm B_R} > 0$) with the lower boundary of $\phiB=10^{\degree}$. While earlier studies have shown that the emissions prefer near-radial magnetic field conditions \citep{malaspina2020}, the current study for the first time quantifies the angular range where waves are observed. It is important to mention that the correlation between the near-$f_{\rm ce}$ harmonics waves and the angular interval estimated in this study is a consistent observation for nearly every such wave event measured by PSP throughout the 11 close encounters with the Sun available until the end of this study (not shown).

\begin{figure}[ht!]
\centering  \includegraphics[width=0.48\linewidth]{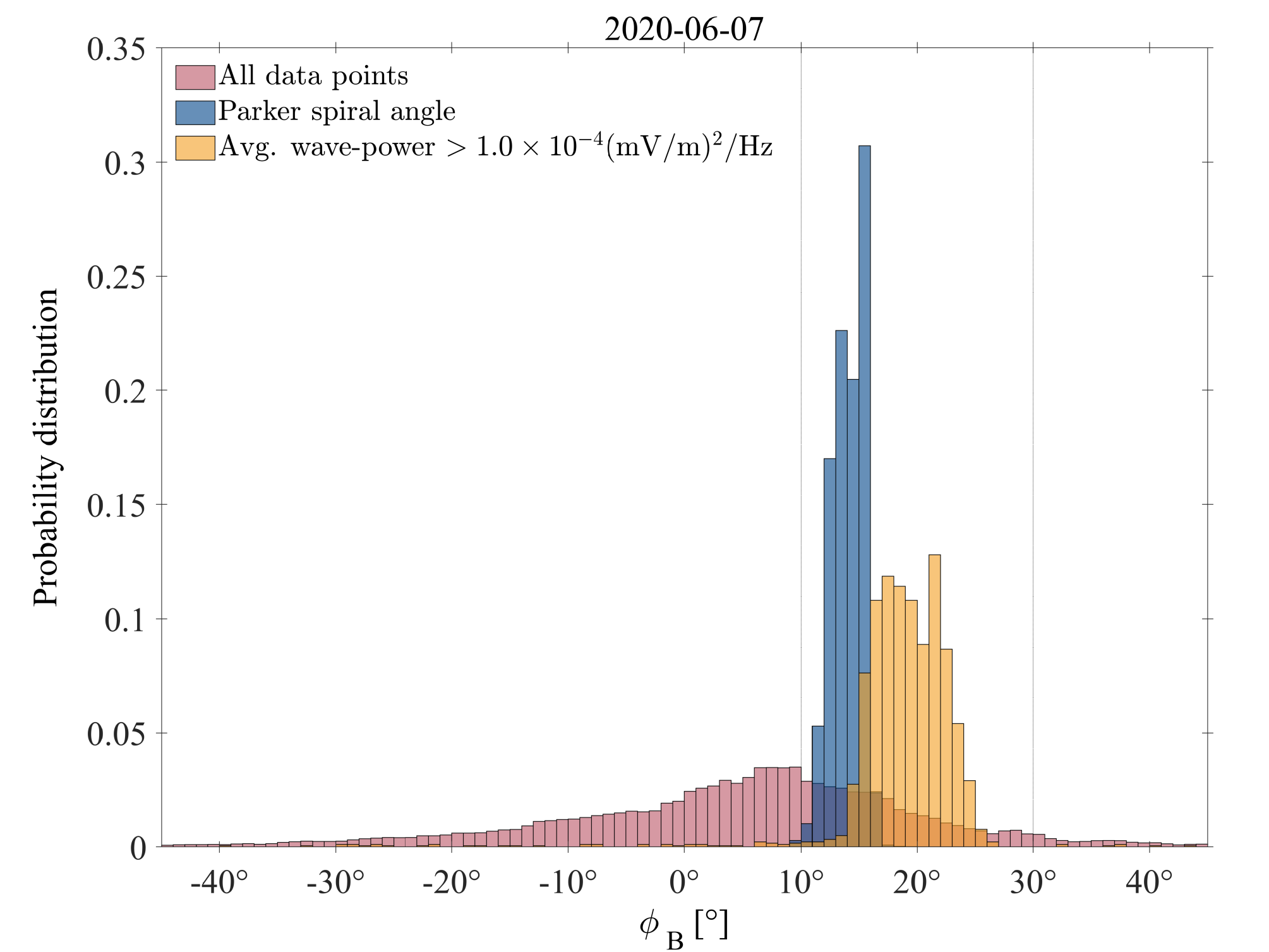}
\centering  \includegraphics[width=0.48\linewidth]{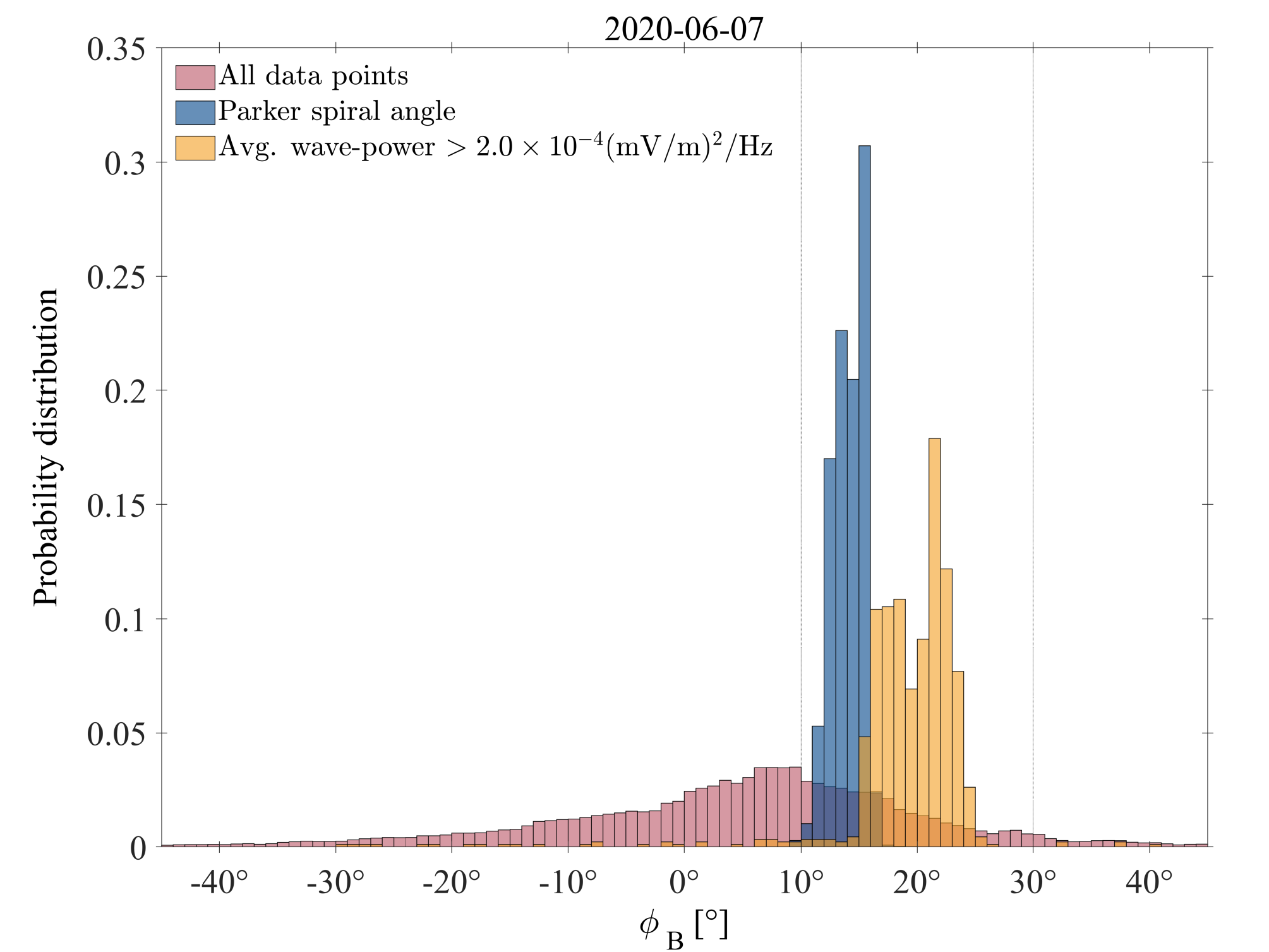}
\caption{Probability distribution with respect to $\phi_{\bf B}$ for the day around the $5^{\rm th}$ orbit perihelion. (red) all data points, (yellow) data points with wave emission power being above (a) $1.0\times 10^{-4}\,\unit{(mV/m)^{2}Hz^{-1}}$ and (b) $2.0\times 10^{-4}\,\unit{(mV/m)^{2}Hz^{-1}}$ , (blue) estimated Parker spiral angle for all data. Each histogram bin corresponds to $1^{\degree}$, and the sum of the bar heights of each quantity in the total angle range ($\pm 90^{\degree}$) is $\leqslant 1$.}
\label{fig:enc5-hist}
\end{figure}

To characterize further the properties of the wave emissions we analyze the probability distribution of wave emission intervals with respect to \phiB. \Cref{fig:enc5-hist} shows three probability distributions for the day around the $5^{\rm th}$ orbit perihelion: the distribution of measured \phiB\ values for all data points (red) and for the data points with wave emissions above a given threshold value (yellow), as well as the estimated values of Parker spiral angle $\phi_{\rm PS}$ for all data points. The probability distribution of data points with wave emissions is shown for two different threshold values: $1.0\times 10^{-4}\,\unit{(mV/m)^{2}Hz^{-1}}$ (on the left) and $2.0\times 10^{-4}\,\unit{(mV/m)^{2}Hz^{-1}}$ (on the right). The vertical dotted lines mark the \phiB\ boundaries that we identify as favorable to wave emission. During this day the average magnetic field has a \phiB\ value slightly below $10^{\degree}$ and the distribution of \phiB\ values for all data points has a broad peak between $-40^{\degree}$ and $+40^{\degree}$. This can be compared to the expected Parker spiral angle values that are in the range of $+11^{\degree}$ to $+17^{\degree}$. We see that on average the magnetic field is more radial than the expected Parker spiral angle. The result is different for magnetic field lines containing the wave emissions. As expected, most of the wave power (yellow bars) is within the dotted lines that mark the \phiB\ angle range favorable to wave emission, with most of the wave power concentrated between $15^{\degree} \lesssim \phiB \lesssim 25^{\degree}$. Thus, during wave emissions, the magnetic field has \phiB\ values that are above the average for the same period. In addition, while the Parker spiral angle distribution  is inside the angular range favorable to wave emission, it is right below the angle interval where the highest wave-power data are distributed. To summarize, the magnetic field during wave emissions has the average value of \phiB\ larger than the average values of magnetic field throughout the interval and larger than the average values of Parker spiral angle. 

Both \thetaB\ and \phiB\ directions are important and have equivalent influence on the emission and properties of near-$f_{\rm ce}$ harmonics waves. Wave emission occurs as soon as \phiB\ and \thetaB\ are simultaneously inside their respective ranges and immediately ceases when at least one of the angles leaves its wave emission interval. Furthermore, the amplitudes of near-$f_{\rm ce}$ harmonics waves increase gradually as both spherical angles approach the center of their respective wave emission intervals. From Figures \ref{fig:scatter-plot} and \ref{fig:enc5-hist}, one can already infer that the highest wave powers tend to occur in an even more restricted angular range, closer to the center of the angle intervals where wave emission occurs.

Another important finding is the correlation between the intensity of near-$f_{\rm ce}$ harmonics and the intensity of broadband magnetic fluctuations within the frequency range between about $2\,\unit{Hz}$ and $200\,\unit{Hz}$. It is known that near-$f_{\rm ce}$ harmonics tend to occur when the magnetic field turbulence at low frequencies is weak. \cite{malaspina2020} shows that statistically more wave emissions are observed in regions of lower magnetic turbulence amplitude at frequencies below $1\,\unit{Hz}$ (which is below proton gyrofrequency). The present study shows, for the first time, that the amplitude decrease of broadband magnetic fluctuations at higher frequencies, well above proton gyrofrequency, is well correlated to the amplitude increase of the near-$f_{\rm ce}$ harmonics. This observation is a consistent feature of these waves, which may suggest a causal relationship between the drop in the magnetic field turbulent energy and the amplitude increase of the high-frequency electrostatic waves.

\begin{figure}[ht!]
\centering  \includegraphics[width=\linewidth]{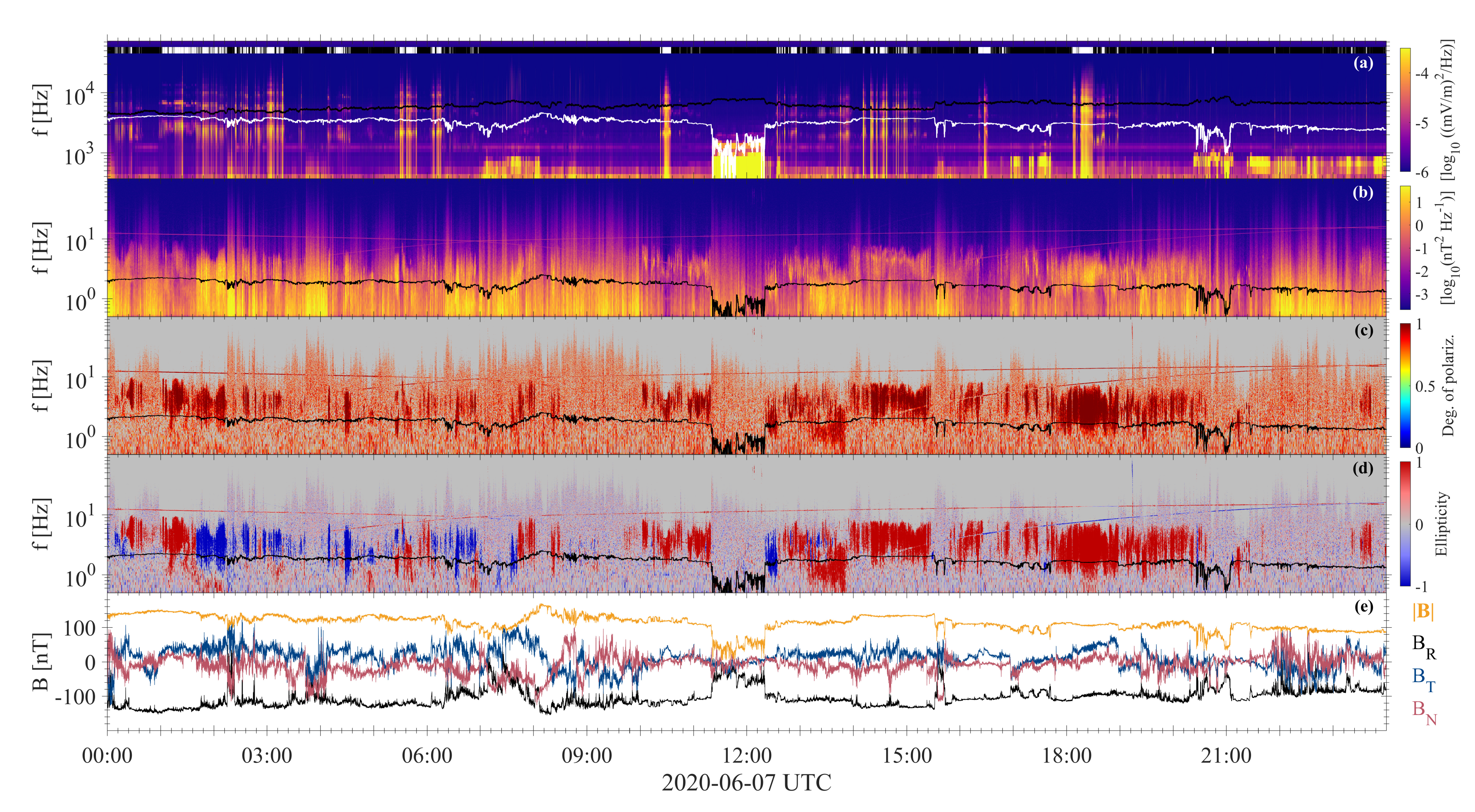}
\caption{High and low-frequency wave emissions around the $5^{\rm th}$ perihelion. (a) High-frequency electric field spectrum from AC differential voltage data measured by ${\rm V}_{12}$ and ${\rm V}_{34}$ antennae pairs (same as \Cref{fig:enc5}(f)). (b) Low-frequency magnetic field spectrogram from fluxgate magnetometer data. (c) Degree of polarization, magnetic field. (d) Ellipticity of the magnetic field waves in the spacecraft frame. Ellipticity is +1 for right-hand and -1 for left-hand polarization. In panels (c) and (d) the gray area corresponds to points where either the degree of polarization is below 0.7 or the wave power  is below $1 \times 10^{-3}\,\unit{nT^2Hz^{-1}}$. In panels (b), (c) and (d) the black line represents the ion gyrofrequency $f_{\rm ci}$. (e) Magnetic field in RTN coordinates, same as \Cref{fig:enc5}(c).
\label{fig:enc5-Bpol}}
\end{figure}

To better understand the relationship between the near-$f_{\rm ce}$ harmonics waves and low frequency waves, we analyze the high-frequency electric field spectrum and the  low frequency waves as observed in magnetic field during encounter 5, see \Cref{fig:enc5-Bpol}. Magnetic field spectrum in  \Cref{fig:enc5-Bpol}(b) shows that in addition to broadband magnetic field fluctuations there are localized wave emissions with frequencies close to and above the proton gyrofrequency (black line). These waves have high degree of polarization, see \Cref{fig:enc5-Bpol}(c), and are left- or right-hand polarized, see \Cref{fig:enc5-Bpol}(d). There is no clear correlation between the intervals of the low-frequency waves and the intervals of the near-$f_{\rm ce}$ harmonics emissions. For instance, in the interval from 07:00 to 10:00, there are no near-$f_{\rm ce}$ harmonics, but there are low-frequency waves with the same polarization properties as the waves observed during near-$f_{\rm ce}$ harmonics events. As an opposite example, there is a near-$f_{\rm ce}$ harmonics event from 16:25 to 16:35 that has no low frequency waves occurring at the same time. Analyzing PSP data from different orbits (not shown here), we notice the same pattern as described above: near-$f_{\rm ce}$ harmonics waves and low-frequency electromagnetic waves are most of the time observed to occur simultaneously, but there are a few instances where both waves are observed to occur independently of each other. Considering these observations, we cannot rule out the possibility of a simple coincidence. Near radial, quiescent magnetic field, with relatively low levels of turbulent fluctuations is a favorable environment for the occurrence of both the high-frequency electrostatic waves with frequencies around $f_{\rm ce}$, and the low-frequency waves with frequencies close to $f_{\rm cp}$, so they end up being observed during the same time intervals most of the time. Unlike the emission of near-$f_{\rm ce}$ harmonics, the emission of low-frequency waves does not have any constraints regarding the magnetic field direction and are observed also at large heliocentric distances. For example, Solar Orbiter has observed these waves at distances above $>40\,\unit{R_{\odot}}$ \citep{khotyaintsev2021}. This clearly shows that the emission of ion-scale waves is independent of the occurrence near-$f_{\rm ce}$ harmonics. However, there is no conclusive evidence showing the opposite situation, i.e., that the emission of near-$f_{\rm ce}$ harmonics is independent of the occurrence of ion-scale waves. The high rate of co-occurrence between the waves and the fact that some of the ion-scale waves can affect near-$f_{\rm ce}$ harmonics properties at small timescales, shows that the waves are not completely uncorrelated. It is possible that the instability exciting the ion-scale waves may be responsible to trigger a secondary instability at electron scales that could lead to the emission of the near-$f_{\rm ce}$ harmonics. In this case, the few instances where we see the near-$f_{\rm ce}$ harmonics waves in the absence of the low-frequency waves could be explained by the stabilization of the ion-scale instability, while the triggered electron-scale instability remains ongoing.

In this study, similar to earlier studies, we could not identify the generation mechanism of the near-$f_{\rm ce}$ harmonics waves. Different mechanisms have been suggested: electron cyclotron drift instability, loss-cone instability, electromagnetic pump wave, and wave-wave interaction \citep{malaspina2021}. Given the observations reported in this study, that the waves favor a particular range of magnetic field directions, do not support any of the previously proposed wave generation mechanisms. Rather the opposite, it is not obvious why the presence of gradients or loss-cone would favor a particular range of magnetic field directions. Similarly, there are no observations that other types of waves would favor similar dependence on the magnetic field. As a result, the possibility that wave-wave interaction is the source of the waves is small. One possibility could be that the emissions are caused by spacecraft interaction with the solar wind, and the direction bias would be due to the geometry of the spacecraft and/or electric field instrument. However, this hypothesis does not explain why the waves are observed only when the broadband magnetic turbulence is low unless their presence somehow affects measurements made by both MAG and SCM instruments. 

Regarding particle analysis, we extensively analyzed the electron pitch-angle distribution at different energy levels (not shown). No evident signature could be consistently related to the near-$f_{\rm ce}$ harmonics or the magnetic field orientation range associated with the emission of these waves. The absence of an identifiable correlation between the electron pitch-angle distribution and the near-$f_{\rm ce}$ harmonics suggests that these waves are generated by electrons with energies outside the energy measurement capabilities of the SWEAP's SPAN-e electrostatic analyzer ($2~\unit{eV}-30~\unit{keV}$) \citep{whittleseySolarProbeANalyzers2020}, which is why we have not included particle analysis in this work. According to the Doppler shift analysis estimates reported in \cite{malaspina2021}, the near-$f_{\rm ce}$ harmonics resonate with electrons with energies between $0.025~\unit{eV}-0.75~\unit{eV}$ for Landau resonance, or $0.2~\unit{eV}-0.8~\unit{eV}$ for N=2 cyclotron resonance. Both energy ranges are way below the lower energy threshold of $2~\unit{eV}$ of the SPAN-e instrument.

So far, the only definitive conclusion we can take from the results presented in this work is that the near-$f_{\rm ce}$ harmonics must be generated locally since their emission and amplitude levels correlate with the local properties of the ambient magnetic field. Any generation mechanism proposed for the near-$f_{\rm ce}$ harmonics must take into account their dependence on the magnetic field direction. Understanding how and why these waves have such a close connection with the magnetic field, including their correlation with low levels of broadband magnetic turbulence, is crucial to identifying the source of these waves and determining whether or not they are spacecraft generated. In a follow-up study (Malaspina et al.~(in preparation)), we provide strong evidence that the near-$f_{\rm ce}$ harmonics are a byproduct of spacecraft interaction with solar wind particles. The basic premise of the study under preparation is that the spacecraft creates an electron shadow region where very low-energy electrons traveling along the background magnetic field cannot access. The waves are emitted when the electron shadow becomes aligned with the ion-wake created by the spacecraft movement. This alignment occurs when the magnetic field is within the range of directions reported in the present work.
\section{Conclusion}
We have used Parker Solar Probe data to investigate the connection between the ambient magnetic field and the emission of near-$f_{\rm ce}$ harmonics waves in the near-Sun solar wind. The results suggest that the magnetic field vector direction is a determining factor for the emission of near-$f_{\rm ce}$ harmonics. Wave emission occurs only when the ambient magnetic field points to a narrow region in space, bounded by polar and azimuthal angular ranges in the RTN coordinate system of correspondingly $80^{\degree} \lesssim \thetaB \lesssim 100^{\degree}$ and $10^{\degree} \lesssim \phiB \lesssim 30^{\degree}$. This dependence on the magnetic field direction is consistent for almost every near-$f_{\rm ce}$ harmonics event measured by PSP throughout its eleven close orbits to the Sun. The few outliers do not deviate more than $5^{\degree}$ from one of the angular boundaries described above. Wave amplitudes are highest when both angles are close to the center of their respective angular interval favorable to wave emissions. In addition, the data shows that the angular range for wave emission, as well as the intensity peaking when both spherical angles are closer to center of their respective wave emission interval, is observed for variations in the magnetic field direction at time scales that vary from large-scale variations on a minute scale down to small-scale magnetic field oscillations at a sub-second scale. 
Near-$f_{\rm ce}$ harmonics emissions are well correlated to a noticeable decrease in the intensity of broadband magnetic field fluctuations in the frequency range $10\,\unit{Hz} \sim 100\,\unit{Hz}$. The higher the amplitude of the near-$f_{\rm ce}$ harmonics, the more pronounced the decrease observed in the magnetic field turbulent spectrum. During the wave emissions, there can be present low-frequency waves with frequency $2\,\unit{Hz}$ to $10\,\unit{Hz}$, which corresponds to a few times the proton gyrofrequency. The low-frequency waves show a high degree of polarization and can be left- or right-handed polarized, while broadband fluctuations do not show a distinct polarization pattern. We could not identify a clear correlation between the near-$f_{\rm ce}$ harmonics and the low-frequency waves. However, due to the high co-occurrence rate between the high and low frequency waves, and the fact that the ion-scale waves may interact with the near-$f_{\rm ce}$ harmonics at small time scales, we could not fully discard a possible correlation. More studies are required to determine if the low-frequency waves actually have any contribution to the emission of the high-frequency harmonics.

Regarding the generation mechanism, the observations presented in this work are inconclusive and do not favor any of the emission mechanisms proposed in \cite{malaspina2021}. The present report on dependence on the magnetic field direction for the emission of the near-$f_{\rm ce}$ harmonics is, to our knowledge, an unprecedented observation in the free solar wind plasma, which suggests that these waves could be related to a distinctive, underlying property of the near-Sun environment. However, even though the idea of discovering a new property of the near-Sun plasma that would cause the magnetic field direction dependence we observe with the near-$f_{\rm ce}$ harmonics is appealing, that is probably not the case. In a follow-up study (Malaspina et al.~(in preparation)), we provide strong evidence that the near-$f_{\rm ce}$ harmonics are generated by the interaction between the spacecraft and the solar wind. Still, the exact generation mechanism of these waves is not entirely understood and requires further investigation.

\begin{acknowledgements}
The authors thank the \emph{Parker Solar Probe}, FIELDS and SWEAP teams. The FIELDS experiment on \emph{Parker Solar Probe} spacecraft was designed and developed under NASA contract NNN06AA01C. All data used in this work are publicly available on the FIELDS (\url{http://research.ssl.berkeley.edu/data/spp/data/}) and SWEAP (\url{http://sweap.cfa.harvard.edu/pub/data/sci/sweap/}) data archives. Data analysis was performed using the IRFU-Matlab analysis package available at \url{https://github.com/irfu/irfu-matlab/tree/PSPdevel}. SFT and AV acknowledge support from Swedish National Space Board Contract 163/19. SFT acknowledges Henriette Trollvik for the helpful conversations regarding the statistical analysis and data visualization and Vin\'{i}cius Ferreira for the insightful discussion about wave modulation.
\end{acknowledgements}

\bibliographystyle{aasjournal}
\bibliography{solarwind}

\begin{thebibliography}{}
\expandafter\ifx\csname natexlab\endcsname\relax\def\natexlab#1{#1}\fi
\providecommand{\url}[1]{\href{#1}{#1}}
\providecommand{\dodoi}[1]{doi:~\href{http://doi.org/#1}{\nolinkurl{#1}}}
\providecommand{\doeprint}[1]{\href{http://ascl.net/#1}{\nolinkurl{http://ascl.net/#1}}}
\providecommand{\doarXiv}[1]{\href{https://arxiv.org/abs/#1}{\nolinkurl{https://arxiv.org/abs/#1}}}

\bibitem[{Bale {et~al.}(2016)Bale, Goetz, Harvey, Turin, Bonnell,
  {Dudok~de~Wit}, Ergun, MacDowall, Pulupa, Andre, Bolton, Bougeret, Bowen,
  Burgess, Cattell, Chandran, Chaston, Chen, Choi, Connerney, Cranmer,
  {Diaz-Aguado}, Donakowski, Drake, Farrell, Fergeau, Fermin, Fischer, Fox,
  Glaser, Goldstein, Gordon, Hanson, Harris, Hayes, Hinze, Hollweg, Horbury,
  Howard, Hoxie, Jannet, Karlsson, Kasper, Kellogg, Kien, Klimchuk,
  Krasnoselskikh, Krucker, Lynch, Maksimovic, Malaspina, Marker, Martin,
  {Martinez-Oliveros}, McCauley, McComas, McDonald, {Meyer-Vernet}, Moncuquet,
  Monson, Mozer, Murphy, Odom, Oliverson, Olson, Parker, Pankow, Phan,
  Quataert, Quinn, Ruplin, Salem, Seitz, Sheppard, Siy, Stevens, Summers,
  Szabo, Timofeeva, Vaivads, Velli, Yehle, Werthimer, \& Wygant}]{bale2016}
Bale, S.~D., Goetz, K., Harvey, P.~R., {et~al.} 2016, Space Science Reviews,
  204, 49, \dodoi{10.1007/s11214-016-0244-5}

\bibitem[{Fox {et~al.}(2016)Fox, Velli, Bale, Decker, Driesman, Howard, Kasper,
  Kinnison, Kusterer, Lario, Lockwood, McComas, Raouafi, \& Szabo}]{fox2016}
Fox, N.~J., Velli, M.~C., Bale, S.~D., {et~al.} 2016, Space Science Reviews,
  204, 7, \dodoi{10.1007/s11214-015-0211-6}

\bibitem[{Kasper {et~al.}(2016)Kasper, Abiad, Austin, {Balat-Pichelin}, Bale,
  Belcher, Berg, Bergner, Berthomier, Bookbinder, Brodu, Caldwell, Case,
  Chandran, Cheimets, Cirtain, Cranmer, Curtis, Daigneau, Dalton, Dasgupta,
  DeTomaso, {Diaz-Aguado}, Djordjevic, Donaskowski, Effinger, Florinski, Fox,
  Freeman, Gallagher, Gary, Gauron, Gates, Goldstein, Golub, Gordon, Gurnee,
  Guth, Halekas, Hatch, Heerikuisen, Ho, Hu, Johnson, Jordan, Korreck, Larson,
  Lazarus, Li, Livi, Ludlam, Maksimovic, McFadden, Marchant, Maruca, McComas,
  Messina, Mercer, Park, Peddie, Pogorelov, Reinhart, Richardson, Robinson,
  Rosen, Skoug, Slagle, Steinberg, Stevens, Szabo, Taylor, Tiu, Turin, Velli,
  Webb, Whittlesey, Wright, Wu, \& Zank}]{kasper2016}
Kasper, J.~C., Abiad, R., Austin, G., {et~al.} 2016, Space Science Reviews,
  204, 131, \dodoi{10.1007/s11214-015-0206-3}

\bibitem[{Khotyaintsev {et~al.}(2021)Khotyaintsev, Graham, Vaivads, Steinvall,
  Edberg, Eriksson, Johansson, {Sorriso-Valvo}, Maksimovic, Bale, Chust,
  Krasnoselskikh, Kretzschmar, Lorf{\`e}vre, Plettemeier, Sou{\v c}ek, Steller,
  {\v S}tver{\'a}k, Tr{\'a}vn{\'i}{\v c}ek, Vecchio, Horbury, O'Brien, Evans,
  \& Angelini}]{khotyaintsev2021}
Khotyaintsev, Y.~V., Graham, D.~B., Vaivads, A., {et~al.} 2021, Astronomy \&
  Astrophysics, 656, A19, \dodoi{10.1051/0004-6361/202140936}

\bibitem[{Livi {et~al.}(2021)Livi, Larson, Kasper, Abiad, Case, Klein, Curtis,
  Dalton, Stevens, Korreck, Ho, Robinson, Tiu, Whittlesey, Verniero, Halekas,
  Mcfadden, Marckwordt, Slagle, Abatcha, \& Rahmati}]{livi2021}
Livi, R., Larson, D.~E., Kasper, J.~C., {et~al.} 2021, Earth Space Sci. Open
  Arch., \dodoi{10.1002/essoar.10508651.1}

\bibitem[{Ma {et~al.}(2021)Ma, Gao, Yang, Tsurutani, Liu, Lu, \& Wang}]{ma2021}
Ma, J., Gao, X., Yang, Z., {et~al.} 2021, The Astrophysical Journal, 918, 26,
  \dodoi{10.3847/1538-4357/ac0ef4}

\bibitem[{Malaspina {et~al.}(2016)Malaspina, Ergun, Bolton, Kien, Summers,
  Stevens, Yehle, Karlsson, Hoxie, Bale, \& Goetz}]{malaspina2016}
Malaspina, D.~M., Ergun, R.~E., Bolton, M., {et~al.} 2016, Journal of
  Geophysical Research: Space Physics, 121, 5088, \dodoi{10.1002/2016JA022344}

\bibitem[{Malaspina {et~al.}(2020)Malaspina, Halekas, Ber{\v c}i{\v c}, Larson,
  Whittlesey, Bale, Bonnell, de~Wit, Ergun, Howes, Goetz, Goodrich, Harvey,
  MacDowall, Pulupa, Case, Kasper, Korreck, Livi, \& Stevens}]{malaspina2020}
Malaspina, D.~M., Halekas, J., Ber{\v c}i{\v c}, L., {et~al.} 2020, The
  Astrophysical Journal Supplement Series, 246, 21,
  \dodoi{10.3847/1538-4365/ab4c3b}

\bibitem[{Malaspina {et~al.}(2021)Malaspina, Iii, Ergun, Bale, Bonnell,
  Goodrich, Goetz, Harvey, MacDowall, Pulupa, Halekas, Case, Kasper, Larson,
  Stevens, \& Whittlesey}]{malaspina2021}
Malaspina, D.~M., Iii, L. B.~W., Ergun, R.~E., {et~al.} 2021, Astronomy \&
  Astrophysics, 650, A97, \dodoi{10.1051/0004-6361/202140449}

\bibitem[{Moncuquet {et~al.}(2020)Moncuquet, {Meyer-Vernet}, Issautier, Pulupa,
  Bonnell, Bale, {de Wit}, Goetz, Griton, Harvey, MacDowall, Maksimovic, \&
  Malaspina}]{moncuquet2020}
Moncuquet, M., {Meyer-Vernet}, N., Issautier, K., {et~al.} 2020, The
  Astrophysical Journal Supplement Series, 246, 44,
  \dodoi{10.3847/1538-4365/ab5a84}

\bibitem[{Mozer {et~al.}(2020)Mozer, Agapitov, Bale, Bonnell, Bowen, \&
  Vasko}]{mozer2020}
Mozer, F.~S., Agapitov, O.~V., Bale, S.~D., {et~al.} 2020, Journal of
  Geophysical Research: Space Physics, 125, e2020JA027980,
  \dodoi{10.1029/2020JA027980}

\bibitem[{Shi {et~al.}(2022)Shi, Zhao, Malaspina, Bale, Dong, Wang, \&
  Wu}]{shi2022}
Shi, C., Zhao, J., Malaspina, D.~M., {et~al.} 2022, The Astrophysical Journal
  Letters, 926, L3, \dodoi{10.3847/2041-8213/ac4d37}

\bibitem[{Whittlesey {et~al.}(2020)Whittlesey, Larson, Kasper, Halekas,
  Abatcha, Abiad, Berthomier, Case, Chen, Curtis, Dalton, Klein, Korreck, Livi,
  Ludlam, Marckwordt, Rahmati, Robinson, Slagle, Stevens, Tiu, \&
  Verniero}]{whittleseySolarProbeANalyzers2020}
Whittlesey, P.~L., Larson, D.~E., Kasper, J.~C., {et~al.} 2020, The
  Astrophysical Journal Supplement Series, 246, 74,
  \dodoi{10.3847/1538-4365/ab7370}

\end{thebibliography}

\end{document}